\documentstyle[12pt]{article}
\date{}
\def\be{\begin{equation}}
\def\ee{\end{equation}}
\def\ba{\begin{eqnarray}}
\def\ea{\end{eqnarray}}
\def\bb{\begin{thebibliography}{999}}
\def\bi{\bibitem}
\def\eb{\end{thebibliography}}
\title{Statistical Distance For Chaotic Maps}
\author{{Ramandeep S. Johal,}\thanks{Postal Address: 1110, 36-C,
Chandigarh -160014, India}
\\
{\it Department of Physics, Panjab University,} \\
{\it Chandigarh -166014, India. }\\e-mail: raman\%phys@puniv.chd.nic.in}
\begin{document}
\maketitle
\begin{abstract}
The purpose of this letter is to define a distance on the underlying
phase space of a chaotic map, based on 
natural invariant density of the map. It is observed  that for logistic map this distance is 
equivalent to Wootters' statistical distance. 
This distance becomes  the Euclidean distance for a map with constant
invariant density.
\end{abstract}
\newpage
Distance between two states of a system, defined in its space of states, is
an important geometric concept. Actually, various kinds of distances (may not
be metrics) exist in literature depending on need and utility. As early as 1922,
Fisher \cite{f22} defined a distance function also called genetic drift, in population 
genetics studies. Wootters' distance \cite{W81} can be defined in a probability space
as well as in the ray space of quantum states. Its extension to density matrices
was achieved in \cite{Br94}. Metrics based on information theory \cite{K59}, which 
distinguish probability distributions have also been defined. Recently, Monge 
distance was defined between Husimi distributions \cite{Ka98}. 

These distances serve different purposes. They can be used to assess the
accuracy of various approximation techniques \cite{Rv97}. A distinguishability 
metric also serves in quantum measurements, which detect weak signals \cite{Br94}. 
Components of a metric tensor can be given definite interpretation in terms
of uncertainties and correlations of operators generating evolution of quantum
states \cite{Ab93}. Some distances possess a somewhat natural property that they 
are equivalent to the Euclidean distance between the respective states.

It appears that more attention has been given to defining distances for
quantum state space. In the classical phase space, the natural distance 
is generally considered to be the Euclidean one. In this letter, I propose 
a definition for  distance in classical phase space underlying a chaotic map.
For simplicity, we confine ourselves to one-dimensional maps. As a typical
example, we study logistic map $x_{n+1} =rx_n(1-x_n)$, which is fully 
chaotic for $r=4$ in the phase space interval [0:1]. This distance will
have close similarity to the Wootters' distance,  although  it is determined
solely by the dynamics of the map.

For our purpose, it is useful to review Wootters' idea of statistical 
distance on a probability space. It is sufficient to consider a system
(e.g. coin) with only two possible outcomes. The probability space is then
one-dimensional. Wootters defined the distance between two probabilties
$p_1$ and $p_2$ as follows:
\be
d(p_1,p_2) = lim_{n\to \infty}\frac{1}{\sqrt{n}} 
 \begin{array}{l}\{\hbox{Maximum No. of
distinguishable probabilites}\nonumber\\
\hbox{between $p_1$ and $p_2$ for $n$ trials}
\} \end{array}. \label{eq1}
\ee
The final expression is then given by
\be
d(p_1,p_2) = \frac{1}{\sqrt{n}}\int^{p_2}_{p_1} \frac{dp}{2\Delta p}
= \int^{p_2}_{p_1}\frac{dp}{2\sqrt{p(1-p)}}, 
\label{eq2}
\ee
where $\Delta p= \sqrt{\frac{p(1-p)}{n}}$ is the root mean square deviation
in the value of $p$ in finite number of $n$ trials. Thus we obtain
\be
d\;(p_1,p_2) = cos^{-1} (\sqrt{p_1p_2} + \sqrt{(1-p_1)(1-p_2)}\;).
\label{eq3}
\ee
Note that this distance is inherently not same as Euclidean distance. However,
the transformation $p=sin^2\;\theta$ allows us to write $d(\theta_1,\theta_2)
= (\theta_2-\theta_1)$. Thus statistical distance becomes equivalent to Euclidean 
distance (angle). 

Now we consider the case of chaotic maps. For such maps, it is known that
the trajectory moves over a strange attractor which can have fractal 
dimension. In the usual box-counting algorithm, to assess the fractal
or capacity dimension, we partition the phase space into boxes or
intervals. We then count the number of times a particular interval is
visted by the map. For an attractor with fractal dimension, the number of intervals
with certain probability show power-law behaviour as the size of the interval
is taken to zero. This helps us to determine the fractal dimension, which is 
intrisically a property of the attractor, independent of the dimension
of the phase space. Note however, that the intervals or boxes are
{\it a priori} defined to be Euclidean or their dimension is the same
as that of the embedding phase space. Now as the capacity dimension
measures how densely a chaotic trajectory covers the phase space, similarly
one can define a distance or interval in phase space measureable in terms 
the chaotic trajectory covers or fills that interval.

With this motivation, we define a distance for chaotic maps as follows:
\be
d'(x_1,x_2) = lim_{n\to \infty}
\begin{array}{l}\{
\hbox{Probability of visiting the 
interval }\nonumber \\ \hbox{between $x_1$ and $x_2$}\},
\end{array} 
\label{eq4}
\ee
where $n$ is the number of iterates and $x_1$ and $x_2$ are any two 
points on the one-dimensional phase space in which attractor is embedded.

Let us apply this idea to the logistic map $x_{n+1}=4x_n(1-x_n)$. We know an
analytic expression for natural invariant density of this map exists, given by $\rho (x) =
\frac{1}{\pi \sqrt{x(1-x)}}$.
Thus the distance as defined in eq.(\ref{eq4}) can be written as
\ba
d'(x_1,x_2) &=& \int^{x_2}_{x_1}\rho (x)dx \label{eq5a} \\
&=& \frac{1}{\pi}\int^{x_2}_{x_1}
\frac{dx}{\sqrt{x(1-x)}}.
\label{eq5b}
\ea
Note the remarkable similarity of the above expression to Wootters' distance
as given by (\ref{eq2}). Thus the two distances are in fact equivalent.

We discuss a plausible reason for this equivalence. Wootters' distance is obtained essentially
by counting the number of distinguishable states or points between $p_1$ and $p_2$.
On the other hand, the distance I define, measures the probability of 
visiting the interval between corresponding points, which is also the 
relative frequency $\frac{n_i}{n}$, where $i$ stands for the interval $(x_1,x_2)$.
So essentially, this distance also measures the number of times the trajectory
comes into the specific interval. We can be sure that no point in the interval
is visited more than once, because of the chaotic nature of the
trajectory which ensures that same point is visited only after infinite  number
of iterations. So all points $n_i$, visited by the trajectory {\it are} distinguishable in
this sense.

Now as such the distance defined in (\ref{eq4}) is not the same as Euclidean
distance. But applying the transformation $x=sin^2\;\frac{\pi y}{2}$, we
can write $d'(y_1,y_2) = y_2 - y_1$. Note that this transformation
is precisely the one that makes logistic map at $r=4$ topologically conjugate
to binary tent map, whose invariant
density is $\rho (y) =1$. Using this value for invariant density, we
directly arrive at the result that $d'(y_1,y_2)$ is equal to Euclidean distance.
 In fact, any one-dimensional chaotic map with constant invariant density
will give a Euclidean distance (within a multiplicative constant)
 according to definition (\ref{eq5a}). 
It is interesting to see that logistic map with non uniform invariant 
density also gives the same result. Naturally, the trigonometric transformation
between $x$ and $y$ variables plays the key role. 

One can look forward to extending this idea of distance to chaotic maps in 
higher dimensions. But more importantly, the simple example given above for
one-dimensional maps should  provide deeper insight into the relation between
statistical fluctuations and the chaotic nature of the trajectory in the asymptotic limit
($n\to \infty$). Finally, we know the box dimension of a chaotic attractor 
tells us how densely
the trajectory covers the underlying phase space. It is interesting to see
whether in some sense, the distance defined above complemets this role. 
More clear exposition of their relation  will be highly welcome. 

\bb
\bi{f22} R.A. Fisher, Proc. R. Soc. Edinburgh, {\bf 42}, 321 (1922).
\bi{W81} W.K. Wootters, Phys. Rev. D, {\bf 23}, 357 (1981).
\bi{Br94} S.M. Braunstein and C.M. caves, Phys. Rev. Lett. {\bf 72}, 3439 (1994).
\bi{K59} S. Kullback, {\it Information Theory and Statistics}, (J. Wiley,
New York, 1959).
\bi{Ka98} K. Zyczkowski and W. Slomczyn'ski, {\it Monge Distance Between 
Quantum States}, (preprint).
\bi{Rv97} M. Ravicule', M. Casas and A. Plastino, Phys. Rev. A {\bf 55},
1695 (1997) and references therein.
\bi{Ab93} S. Abe, Phys. Rev. A {\bf 48}, 4102 (1993).
\bi{Mac94} J.L. MaCauley, {\it Chaos, Dynamics and Fractals: An algorithmic
approach to deterministic chaos }, Cambridge University Press, 1994, p. 75.
\eb
\end{document}